\documentstyle[preprint,aps]{revtex}
\newcommand{\subs}[1]{{\mbox{\scriptsize #1}}}

\begin{document}
%\preprint{EA-NS 1997-1}

%\draft
\begin{title}
{Torsion and electron motion in Quantum Dots
with crystal lattice dislocations}
\end{title}

\author{E. Aurell}
\address{Department of Mathematics, Stockholm University
          S--106 91 Stockholm, Sweden\\
	PDC/KTH, S-100 44 Stockholm, Sweden}

\date{\today}

\maketitle

%\pacs{61.72Bb, 73.20.Dx, 85.30.Vw}

\begin{abstract}
The motion of a conducting electron in a quantum dot with
one or several dislocations in the underlying crystal lattice is
considered in the continuum picture, where dislocations are
represented by torsion of space.
The possible effects of torsion are investigated on the
levels of classical motion, non-relativistic quantum motion,
and spin-torsion coupling terms derivable in the
non-relativistic limit of generalizations of the
Dirac equation in a space with torsion.
Finally, phenomenological spin-torsion couplings analogous
to Pauli terms are considered in the non-relativistic
equations.
Different prescriptions of classical and non-relativistic
quantum motion in a space with torsion are shown to give
effects that should in principle be observable.
Semi-classical arguments are presented to show that 
torsion is not relevant for the
classical motion of the center of a wave packet.
The correct semi-classical limit can instead be described 
as classical trajectories in a Hamiltonian given by the band energy.
In the special case of a spherically symmetric band this motion
reduces to straight lines, independently of local crystal orientations. 
By dimensional analysis the coupling constants of the
possible 
spin-torsion interactions are postulated to be
proportional to a combination of the effective mass 
of the electron, $m_\subs{eff}$,
the lattice constant, $a$, and $\hbar$.
The level-splitting is then very small
with transition frequencies on the order of 1~kHz or smaller.

\end{abstract}

\section{Introduction}
\label{s:introduction}
Quantum dots are small essentially two-dimensional
 conducting domains, 
 connected to external leads by tunnel barriers
such that the
number of conducting electrons is quantised.
The elastic mean free path and the dephasing length 
may, under appropriate circumstances,
be much larger than  lateral size of the dots.
The motion of an electron inside may therefore to a first approximation 
be considered as ballistic.
These structures
may hence in many respects be regarded as physical
realizations of the quantum mechanics text-book examples of motion in
two-dimensional potential wells. For recent reviews and papers,
see \cite{KirkReed,Kastner,BarangerJalabertStone,TomsovicHeller,Marcus,Mathur,Agam,Bruus}.

The purpose of this Letter is to explore a suggestion made in
a book and in a
series of papers by Kleinert
and co-workers\cite{Kleinert,KleinertFiziev,KleinertLANL},
which I will
state compactly as follows:
i) the motion of a quantum particle in a space with
torsion has several distinct features compared to free motion
in flat space;
ii) a crystal with defects may in the continuum picture be described
as a space with torsion;
iii) the motion of an electron around a defect may thus in some 
circumstances be modelled by the motion of a quantum particle in a space
with torsion.

A continuum description of a crystal
 is valid at distances much larger than the
lattice spacing.
The low-lying electronic states in a quantum dot extend across
the dot, with typical spatial scale $L$, while a localized defect
extends over a few lattice spacings $a$.
The scale separation $L/a$ is thus in the range of $100-10000$,
and it is conceivable that an effective continuum
description might be appropriate.

The outline of this paper is as follows.
In section~\ref{s:classical-quantum} I review the literature
on motion in spaces with torsion.
The Dirac equation coupled to torsion allows for one
vector and one axial vector coupling, both of which give rise
to Zeeman-like couplings of spin to an axial vector in the non-relativistic
limit. We also consider more phenomenologically motivated spin-torsion
couplings which yield similar terms.
In section~\ref{s:torsion} I  review elementary aspects
of the differential geometry of physics of defects
and in section~\ref{s:geodesics-vs-autoparallels} I discuss on
the semi-classical level the effects of crystal torsion on elecron motion. 
In section~\ref{s:defect} I try to evaluate possible torsion-induced
effects in quantum dots.
In section~\ref{s:conclusion} I summarize the results.
They are are negative as
concerns the proposal by Kleinert that autoparallels should play a
privileged role.
On the semi-classical level it is possible to observe motion along
geodesics, but such motion is insensitive to torsion.
Motion along autoparallels and a related non-relativistic quantum
mechanics could also probably be disproved by experiments.
The spin-torsion coupling terms are not theoretically unsound,
but leads to very weak effects that they would 
be difficult to observe.

\section{Classical and quantum motion in a space with torsion}
\label{s:classical-quantum}
In this section I use the Einstein
conventions of summing over repeated indices and raising and
lowering indices by the metric tensor $g_{ij}$ and its inverse $g^{ij}$.
For conventions pertaining to the affine connection, 
torsion and contorsion
I follow  Hehl et al\cite{Hehl}, see also
Schr\"odinger\cite{Schrodinger} and Schouten\cite{Schouten}.

A manifold is said to carry metric and affine structure
if one can compute the length of vectors in the tangent space
by
\begin{equation}
|A|^2 = g_{ij}A^iA^j
\label{eq:vector-length}
\end{equation}
and parallel transport of a vector along an infinitessimal distance
$d\vec x$ by
\begin{equation}
A^i \to A^i - \Gamma_{kj}\,^{i} A^j dx^k
\label{eq:vector-parallel-transport}
\end{equation}
The metric and the affine
connection $\Gamma$ are connected by the compatibility condition,
that the scalar product of two arbitrary
vectors is invariant under parallel
transport:
\begin{equation}
g_{ij,k} - g_{lj}\Gamma_{ki}\,^{l} - g_{il}\Gamma_{kj}\,^{l}  = 0
\label{eq:vector-compatibility-condition}
\end{equation}
which may also regarded as the statement that $g$ is covariantly
constant.
The first term in (\ref{eq:vector-compatibility-condition}) stands for
the partial derivative $\partial g_{ij}/\partial x^k$.

The standard form of the solution of 
(\ref{eq:vector-compatibility-condition})
is
\begin{equation}
g_{il}\Gamma_{kj}\,^{l}  = g_{il}\{ {{l}\atop {kj}} \} - K_{kji}
\qquad \{ {{l}\atop {kj}} \} = \frac{1}{2}g^{lm}
\left(g_{mj,k}+g_{mk,j}-g_{kj,m}\right)
\label{eq:contorsion-definition}
\end{equation}
where we recognize
the Cristoffel
symbol $\{ {{l}\atop {kj}} \}$, 
symmetric under the interchange of $j$ and $k$.
The second term in (\ref{eq:contorsion-definition}), $K_{kji}$,
is known as contorsion and
must be antisymmetric under the interchange of the last two indices
of $j$ and $i$.

Let us consider a small area spanned by two vectors
$dx_{(1)}$ and $dx_{(2)}$. If we transport  $dx_{(2)}$
along $dx_{(1)}$ it will change into (in component form)
$dx^i_{(2)}- \Gamma_{kj}\,^{i}dx^j_{(2)}dx^k_{(1)}$.
If we move first along $dx_{(1)}$ and then along $dx_{(2)}$
we will hence end up with a displacement (in component form)
\begin{equation}
dP_{12}^i = dx^i_{(1)} + dx^i_{(2)}- \Gamma_{kj}\,^{i}dx^j_{(2)}dx^k_{(1)}
\end{equation}
If we were on the other hand to make the displacements in the opposite
order we would end up at
\begin{equation}
dP_{21}^i = dx^i_{(2)} + dx^i_{(1)}- \Gamma_{kj}\,^{i}dx^j_{(1)}dx^k_{(2)}
\end{equation}
The difference between these displacements is a vector, which measures
by how much the circuit fails to close if direction vectors are parallel
transported around the perimeter of the area spanned by 
$dx_{(1)}$ and $dx_{(2)}$
\begin{equation}
dB^i =dP_{12}^i-dP_{21}^i
=  \Gamma_{kj}\,^{i}\left( dx^j_{(1)}dx^k_{(2)}-dx^j_{(2)}dx^k_{(1)} \right)
\label{eq:dB-definition}
\end{equation}
The parenthesis in (\ref{eq:dB-definition}) is the area element
$dA_{12}\,^{jk}$ spanned by $dx_{(1)}$ and $dx_{(2)}$.
The part of $\Gamma$ antisymmetric in interchange of
$j$ and $k$ is called torsion,
and has hence the following relation to
$dB$ (in component form):
\begin{equation}
dB^i = -  S_{kj}\,^{i} dA_{12}\,^{kj} \qquad
S_{kj}\,^{i} = \Gamma_{[kj]}\,^{i}
\label{eq:torsion-definition}
\end{equation}
Torsion is therefore a third-order tensor, and it may be
connected to contorsion by
\begin{equation}
K_{ijk} = -S_{ijk} + S_{jki} - S_{kij}
\label{eq:torsion-contorsion-connection}
\end{equation}
from which it is seen that the required
antisymmetricity of $K$
(in the last two indices) follows from the antisymmetricity of $S$
(in the first two indices).
As a final consistency check one
may antisymmetrize
$\Gamma$ using the decomposition
of~(\ref{eq:contorsion-definition}):
\begin{eqnarray}
\Gamma_{[kj]}\,^{i} &=& 
\, \{ {{i}\atop {[kj]}} \}\, - K_{[kj]}\,^{i} 
= \, 0\, -\frac{1}{2}\left(K_{kj}\,^{i}-K_{jk}\,^{i}\right)\nonumber \\
&=& \frac{1}{2}\left(S_{kj}\,^{i}-S_{jk}\,^{i}\right) = S_{kj}\,^{i}
\label{eq:consistency-check}
\end{eqnarray}
The symmetric part of $\Gamma_{kj}\,^{i}$ under interchange of $k$ and
$j$ thus contains both 
the Cristoffel symbol
and a symmetrized combination of the antisymmetric
part, which explains the distinction between torsion and contorsion.

The geodesics on a manifold are
given by
\begin{equation}
\frac{d^2 x^i}{d\tau^2} +
\{ {{i}\atop {jk}} \} \frac{d x^j}{d\tau}
\frac{d x^k}{d\tau} = 0\qquad\hbox{(geodesics)}
\label{eq:geodesics-definition}
\end{equation}

The intuitive concept of free motion seems however to
imply that velocity vectors change according to the law
of parallel transport, and such curves are called autoparallels:
 \begin{equation}
\frac{d^2 x^i}{d\tau^2} +
\Gamma_{kj}\,^{i}
 \frac{d x^j}{d\tau}	
\frac{d x^k}{d\tau} = 0\qquad\hbox{(autoparallels)}
\label{eq:autoparallels-definition}
\end{equation}
The interesting observation stressed by Kleinert is then that
in a space with torsion geodesics and autoparallels do
not coincide.
It is not obvious which of the two is the most natural
extrapolation from ordinary classical mechanics.
For different points of view, see,
on the side of autoparallels,
Kleinert\cite{Kleinert,KleinertFiziev,KleinertLANL},
and on the the side of
geodesics
Hehl et al\cite{Hehl}, Audretsch\cite{Audretsch},
and the recent papers
of L\"ammerzahl\cite{Lammerzahl} and Barros e S\'a\cite{Nuno}.

In any case, from the viewpoint of Hamiltonian dynamics
autoparallels have peculiar features.
One of the simplest examples is motion on a two-dimensional surface
with constant diagonal metric and constant torsion.
The torsion tensor has then only two independent non-zero components,
namely $S_{121}$ and $S_{122}$, which together specify a direction in
the plane, $\vec b = (S_{121},S_{122})$. It is straightforward to
integrate equations (\ref{eq:autoparallels-definition}): they 
here describe motions
at constant speed, but where the direction of velocity 
changes so as to be orthogonal
to~$\vec b$.
In other words, kinetic energy in the ordinary sense is conserved,
but momentum and phase space volume is not.
In these respects motion along autoparallels
 in a space with torsion is similar to a
mechanical  system with  non-holonomic
constraints\cite{ArnoldKozlovNeishtadt,Poincare}
(for analytic and numerical investigations of
a clarifying concrete example, see \cite{Bondi,GarciaHubbard,Nordmark}),
an analogy also used extensively by Kleinert.

The difference between autoparallels and geodesics carry over
to the quantum mechanics of a non-relativistic scalar particle.
The gradient of a scalar is a vector, and the covariant derivative
of this vector involves the connection, so that
the contraction of two covariant derivatives acting on
a scalar is
\begin{equation}
  g^{ij}D_iD_j\psi =  g^{ij}D_i\partial_j\psi = 
\left(g^{ij}\partial_i\partial_j - g^{ij}\Gamma_{ij}\,^{l}\partial_l
\right)\psi
\label{eq:Laplace-Kleinert}
\end{equation}
The usual generalization of the kinetic term
in the Schr\"odinger equation involves on the other hand the
Laplace-Beltrami operator
\begin{equation}
\frac{1}{\sqrt{g}}\left(\partial_{i}
g^{ij}\sqrt{g}\partial_j\right)\psi
= \left(g^{ij}\partial_i\partial_j - g^{ij}
\{ {{l}\atop {ij}} \} \partial_l
\right)
\psi
\label{eq:Laplace-Beltrami}
\end{equation}
which is self-adjoint with respect to the standard scalar product
$<\eta |\psi>= \int \sqrt{g}\, \eta^*\psi$.

The difference between (\ref{eq:Laplace-Kleinert})
and (\ref{eq:Laplace-Beltrami}) is a gradient
\begin{equation}
K_{i}\,^{il} \partial_l \psi
= -2 S^l\partial_l \psi\qquad S_l = S_{li}\,^i
\label{eq:Beltrami-Kleinert-difference}
\end{equation}
which, as remarked by Kleinert, is not in general self-adjoint,
at least not with respect to the same scalar product.
Quantum dynamics governed by
(\ref{eq:Laplace-Kleinert})
is therefore not necessarily unitary.
This somewhat disturbing
property can be compared with the fact
that phase space volume in the usual sense is not conserved
by classical motion along autoparallels. 

The problem of
geodesics vs.
autoparallels can finally also be considered,
albeit a little indirectly, for relativistic
spin-$\frac{1}{2}$ particles.
To define spinors in General Relativity one needs a system
of local intertial frames $\xi^{\alpha}_X$, each defined
in the neighbourhood of some space-time point $X$.
The transformation matrix between $\xi^{\alpha}_X$
and a chosen, in general
non-inertial local coordinate system at $X$ is called
a vierbien
\begin{equation}
V^{\alpha}\,_{\mu} = \left(\frac{\partial\xi^{\alpha}_X(x)}
{ \partial x^{\mu} }\right)_{x=X}
\label{eq:vierbien-definition}
\end{equation}
The first index of the vierbein stands for a direction in the
local inertial frame, and is hence lowered and raised with the
Minkowski metric $\eta$.
The second index stands on the other hand for a direction in the
tangent space at $X$, in a basis determined by the coordinate system
$x^{\mu}$, and is therefore raised and lowered with the metric
tensor $g$.

Parallel transport and covariant differentiation of a spinor
in direction $\alpha$ of the local inertial frame are
determined by the vierbiens as
\begin{equation}
{\cal D}_{\alpha}
\psi = V_{\alpha}\,^{\mu} \left(
\partial_{\mu} + \Gamma_{\mu}\right)\psi
\qquad \Gamma_{\mu} =\frac{1}{2} \sigma^{\beta\gamma}
 V_{\beta}\,^{\nu}  V_{\gamma\nu;\mu}
\label{eq:covariant-derivative-on-spinor}
\end{equation}
where $\sigma^{\beta\gamma}$ are the basis elements of
infinitessimal
Lorentz transformations in the spinor
representation, and the semicolon in the last
factor stands for covariant differentiation of the vierbien
according to
\begin{equation}
V_{\gamma\nu;\mu} =  V_{\gamma\nu,\mu} - 
\{ { {\kappa}\atop {\mu\nu} } \} V_{\gamma\kappa}
\label{eq:covariant-derivative-on-vierbein}
\end{equation}
The Dirac equation in a space with curvature is then
\begin{equation}
i\hbar\gamma^{\alpha}{\cal D}_{\alpha}\psi + mc {\bf 1}\psi = 0
\label{eq:Dirac-equation-curvature}
\end{equation}
For a classical discussion of these matters,
see \cite{Weinberg}.
For a thoroughly modern discussion from the mathematical point
of view, see \cite{BerlineGetzlerVergne}.

When we allow also torsion of space we enter into less well-chartered
territory.
One possible way to proceed is to take as building blocks
the vierbeins directly, allowing for
$V^{\alpha}\,_{\mu,\nu}$ not necessarily being equal to 
$V^{\alpha}\,_{\nu,\mu}$.
The minimal formal
change in the Dirac equation
is then to perform the covariant
differentiation in (\ref{eq:covariant-derivative-on-vierbein}) using
the full affine connection instead of the Cristoffel symbol.
However, we have now also torsion as an independent tensor.
By contractions we can form a
vector 
$S^{l}=S^{l}\,_{m}\,^m$ and an axial vector
$\tilde S^{l}=\epsilon^{lmno}S_{mno}$.
The simplest relativistically covariant
additional terms in (\ref{eq:Dirac-equation-curvature}) are hence
\begin{equation}
\left(i  C\hbar S_{\alpha}\gamma^{\alpha} + 
 D\hbar \tilde S_{\alpha}\gamma_5\gamma^{\alpha}\right)\psi
\label{eq:Dirac-equation-torsion-addenda}
\end{equation}
where $ C$ and $ D$ are dimension-less numerical constants.
In fact, doing the covariant differentiation
in (\ref{eq:covariant-derivative-on-vierbein})  with the affine connection
will not bring in any more
terms except those in (\ref{eq:Dirac-equation-torsion-addenda}).
The Dirac equation coupled to torsion by the simplest vector
and axial vector couplings and covariant differentiation 
is thus
\begin{equation}
i\hbar\gamma^{\alpha}{\cal D}_{\alpha}\psi 
+\left((C-1)i\hbar S_{\alpha}\gamma^{\alpha}
+(D+\frac{1}{6})\hbar \tilde S_{\alpha}\gamma_5\gamma^{\alpha}\right)\psi
+mc {\bf 1}\psi = 0
\label{eq:Dirac-equation-torsion}
\end{equation}
where the covariant derivative ${\cal D}_{\alpha}$ is defined
by (\ref{eq:covariant-derivative-on-spinor})
and (\ref{eq:covariant-derivative-on-vierbein}).
If we also put the constraint
that (\ref{eq:Dirac-equation-torsion}) should be derivable
by variation from a real action, we are led to the specific
choice $C=1,D=0$, which is often in the literature
referred to as the Dirac equation minimally   
coupled to torsion\cite{HehlDatta,Hehl,Audretsch,Lammerzahl}.

I will in the following neglect time components of the space-time
torsion vector $S_l$. 
The only non-vanishing component of the axial vector $\tilde S_l$ is
then $\tilde S_0$, which, if we look at it in 3D, transforms as
a pseudo-scalar.
The axial vector coupling in (\ref{eq:Dirac-equation-torsion})
has then the following 
non-relativistic limit \cite{Lammerzahl}
\begin{equation}
i(D+\frac{1}{6})
\frac{\hbar^2}{m}\tilde S_0\left(\vec\sigma\cdot\vec\partial\right)\psi
\label{eq:Lammerzahl-coupling}
\end{equation}
where $\vec\sigma$ is the vector of Pauli matrices and the derivative
acts on the wave function to the right.

The vector coupling in (\ref{eq:Dirac-equation-torsion}) is
on the other hand
completely analogous to the coupling to an external electro\-magnetic
field. It will thus, 
in the non-relativistic limit, give rise to an effect like
the coupling of spin and magnetic field
\begin{equation}
\frac{(C-1)}{2}\frac{\hbar^2}{m} (\vec\nabla\times\vec S)\cdot\vec\sigma\psi
\label{eq:Zeeman-coupling}
\end{equation}
where $\vec S$ stands for the vector
of spatial components of the four-dimensional torsion vector.

In summing up this section we see that the equations for autoparallels
(\ref{eq:autoparallels-definition})
and the putatitive generalization of the Schr\"odinger operator as the
contraction of two covariant derivatives formed with the affine
connection  
(\ref{eq:Laplace-Kleinert})
are the odd ones out.
Neither arises as the limit of the Dirac equation minimally coupled
to torsion, which  only gives a 
spin-torsion coupling reminiscent of a Pauli term,
the standard Laplace-Beltrami
operator and
geodesics.

\section{Torsion: continuum representation of defects}
\label{s:torsion}
The description of defects in crystal lattices in 
the continuum picture with tools from differential geometry
has a distinguished history, described in the
1980 Les Houches lectures of Kr\"oner\cite{Kroner},
which also give a good introduction.

The basic idea is quite simple.
A crystal is supposed to carry only isolated defects,
such that one has around most points a locally
perfect lattice.
The lattice locally
around a point, if continued 
without defects or deformations indefinitely, may be considered as
the tangent space of the crystal at this point.
The local lattice directions provide
a basis for this space.
A vector in the tangent space can then be identified
with moving a certain number of lattice units
in each direction.
Parallel transport of a vector from lattice point $P$
to lattice point $Q$ means that we identify
the vector of, say, $n_i$ steps
in crystal direction $i$ at $P$ with 
the vector of an equal number of steps in the same
lattice direction
at $Q$.
Since the local lattice directions may change from point
to point in the crystal, a vector can change under parallel
transport if measured in an external frame of reference.

We can then consider the following process.
Take two lattice vectors $n_1$ and $n_2$
in directions $i_1$ and $i_2$.
Transport the pair first along  $n_1$, then
along $n_2$, $-n_1$ and $-n_2$.
If the crystal is perfect the circuit closes and we
are back at the point where we started.

If, however, the circuit has circled a dislocation line,
we are not back at the point of departure.
Dislocations lines in three-dimensional crystals
are characterized by a vector $\vec t$ tangential to the line,
and a vector $\vec b$ describing the mismatch if we circle
the dislocation line in the positive sense determined by
$\vec t$.

That is, if we introduce a vector $dA_j$
normal to
and equal in length to the area spanned by $n_1$ and $n_2$,
then the mismatch when going around the
circuit is linearly related to $dA$ by
\begin{equation}
db^i = \alpha^{ji} dA_{j}
\end{equation}
where $\alpha$
is the density of dislocation lines with $\vec t$
in the $j$ direction and $\vec b$ in the $i$ direction\cite{Kroner}.

The procedure described here is of course
identical to
that used to define the torsion tensor in section~\ref{s:classical-quantum}
so one may introduce a crystal torsion field as
\begin{equation}
S_{kj}\,^i = - \frac{1}{2} \epsilon_{kjm} \alpha^{mi} 
\end{equation}
where $\epsilon$ is the totally antisymmetric Levi-Civita tensor
and summation of repeated indices is understood.

In three-dimensional simple
crystals with one atom per Bravais cell
there are two qualitative types of dislocations:
screw dislocations where $\vec b$ is parallel to $\vec t$;
and edge dislocations where $\vec b$ is normal to
$\vec t$\cite{Kroner,AshcroftMermin,Kittel}.
In the first case the dislocation tensor $\alpha$ has only
diagonal elements,
while in the second case it has only off-diagonal elements.

The totally antisymmetric trace of the crystal torsion field vanishes
for edge dislocations since
\begin{equation}
\epsilon^{kji}S_{kji} = -  \alpha^{l}\,_{l}
\end{equation}
On the other hand, for screw dislocations the contracted torsion
vector vanishes since
\begin{equation}
S_i = S_{il}\,^l = -\frac{1}{2}\epsilon_{ilm}\alpha^{ml}
\end{equation}
$S_i$ can hence be looked upon as a
vector dual to the
antisymmetric part of the dislocation density tensor.
For edge dislocations it
is normal to both $\vec b$ and $\vec t$,
and equal to half of $\vec b$ in length.

\section{Geodesics vs. autoparallels in crystal space}
\label{s:geodesics-vs-autoparallels}
A crystal space with dislocations but without interstitial defects
carries torsion but not curvature\cite{Kroner}.
Geodesics in this space means simply
straight lines in the
frame of reference of an external observer.
Autoparallels means on the other hand motion which is always
in the same direction with respect to the local crystal
directions.

The following discussion will 
be on the level of a semiclassical
model.
The basic idea is to take
a wave packet, a superposition of
Bloch states, and assume that
the crystal changes on
length scales much larger than the spread of the
wave packet.
To construct the Bloch states I can therefore take
the crystal lattice close to a given point
where the wave packet is centered and extend
it indefinitely in all direction without deformations
or defects. This 
corresponds to the tangent lattice in the sense
of section~\ref{s:torsion}.
If I go far enough from the point the real lattice and the tangent
lattice will differ, but at those distance the amplitude
of the wave packet is assumed vanishingly small, so this difference
will be ignored.

More quantitatively, the wave packet is built from Bloch states
with wave vectors in a domain of size
$\Delta k$ around  ${\bf k}$.
We assume that $\Delta k$ is small compared to the dimensions of the Brillouin zone,
and that  $1/\Delta k$ is 
small compared to the scale $L$ on which the
lattice changes orientation appreciably.
In real space the wave packet is then
located in a domain of size $1/\Delta k$
around a center, which is denoted $\bf r$.

In the absence of external electric and magnetic fields the equations
of motion for the center of the wave packet in an undistorted crystal
are\cite{AshcroftMermin}
\begin{equation}
\dot{\bf r} = {\bf v}_n({\bf k}) = \frac{1}{\hbar}
\frac{\partial {\cal E}_n({\bf k})}{\partial {\bf k}}
\qquad \hbar \dot {\bf k} = 0
\label{eq:Ashcroft-Mermin-equation}
\end{equation}
where $\hbar{\bf k}$ is the crystal momentum
and 
${\cal E}_n({\bf k})$ is the energy
of the state with wave vector ${\bf k}$ in the $n$'th band.
Equation~(\ref{eq:Ashcroft-Mermin-equation}) describes
motion in a straight line.
This picture is valid as long as dispersion effects are
not important, 
\begin{equation}
t<< t_\subs{disp} \sim \frac{\hbar}{
\frac{\partial^2 {\cal E}_n({\bf k})}{\partial {\bf k}^2}
(\Delta k)^2
}
\label{eq:dispersion-time}
\end{equation}
To see the effects of the changing crystal before dispersion sets
in too strongly we must demand that $L$ is much less than
the distance the wave packet traverses during time $t_\subs{disp}$,
that is
\begin{equation}
\frac{1}{\Delta k} << L << L_\subs{disp}\qquad
L_\subs{disp} = 
\frac{ \frac{\partial {\cal E}_n({\bf k})}{\partial {\bf k}}}
{\frac{\partial^2 {\cal E}_n({\bf k})}{\partial {\bf k}^2}}
\frac {1}{(\Delta k)^2}
\label{eq:dispersion-length}
\end{equation}
As we will see later,
for realistic values of $\Delta k$ the bounds in
(\ref{eq:dispersion-length}) are a little tight, but
for the present discussion it is sufficient that there
are in principle
some scales of time where the wave packet
is still localized on a length scale much less than $L$,
but has moved a distance much larger than $L$.
In this intermediate
regime one can thus pose the problem if the wave packet
would follow geodesics or autoparallels
or some other curves in crystal space.

Let the motion of the center of the wave packet
be parametrized by $(x^*(t),{\bf k}(t))$,
both from now on
given in a coordinate system fixed in space.
The changing orientation of the crystal can then be described
by postulating that the band energy ${\cal E}_n({\bf k},x)$ is
a slowly varying function of $x$.
As the wave packet moves in the crystal from 
position $x^*(0)$ at $0$ to position $x^*(t)$ at $t$ it
acquires a phase of $e^{i\frac{S[x^*(t),{\bf k}(t)]}{\hbar}}$
where
\begin{equation}
S[x^*(t),{\bf k}(t)]
=
\int_0^t [
\hbar {\bf k}(t)\cdot \dot x^*(t)  - {\cal E}_n\left({\bf k}(t),x^*(t)
\right)]
\end{equation}
The actual path is determined by the condition that $S$ should
be stationary under variations.
It is obvious that if we introduce the 
momentum ${\bf p}=\hbar {\bf k}$
and the Hamiltonian function
$H({\bf p},x)= {\cal E}_n(\frac{{\bf p}}{\hbar},x)$
the center will follow the
classical trajectories of $H$.

Let us assume for simplicity
that the band energy has the structure 
${\cal E}_n({\bf k}) = \frac{1}{2}\hbar^2 m_{ij}^{-1} {\bf k}^i {\bf k}^j$,
where $m_{ij}$ is the effective
 mass tensor,
which has principal axes $\hat n^{(1)},\ldots,\hat n^{(d)}$
and effective masses in those directions 
$m_1,\ldots,m_d$.
The changing orientation of the crystal is  effected  by letting
the principal axes depend on $x$.
The classical Hamiltonian is then
\begin{equation}
\sum_i \frac{1}{2m_i} ({\bf p}\cdot \hat n^{(i)}(x))^2
= \sum_{jk} g_{jk}(x) {\bf p}^j {\bf p}^k
\qquad g_{jk}(x) = \sum_i \frac{1}{2m_i}\hat n^{(i)}_j(x)n^{(i)}_k(x)
\end{equation} 
The wave packet will hence
follow the geodesics with respect to the
metric $g_{ij}$ induced by the crystal orientations and the band
structure.

In the special case of a spherically symmetric band structure the
effective mass tensor is proportional to the identity.
The Hamiltonian is then
$\frac{1}{2}m_\subs{eff}^{-1}{\bf p}^2$, and
the dependence of the crystal orientations drop out.
In the last idealised case we would thus predict motion along straight
lines 
which are geodesics and not autoparallels in crystal space.

To end the discussion we must also take into account that lattice
directions in real crystals
are changed by the presence of dislocations.
These are local scatterers, and we have to check if the coherence
of the wave packet can be maintained over such distances that the
lattice orientations change appreciably.
Suppose that the strength of the perturbation is $U$ and that
it has support over a typical distance of the lattice spacing $a$.
The wave packet passes over the defect
during a time $T_\subs{scatter}$ 
of about
$\frac{\hbar}{\Delta k
|\frac{\partial {\cal E}_n({\bf k})}{\partial {\bf k}}| }$, while
the natural time scale of the action of the scattering potential
on the wave packet is $\hbar/U$.
If we assume scattering to any
state on the energy shell with equal probability, we
have
\begin{equation}
P_\subs{scatter on shell} \sim \left(a\Delta k\right)^{d} 
\left(a{\bf k}\right)^{d}
\left(\frac{\Delta k}{\bf k}\right) 
\left(\frac{UT_\subs{scatter}}{\hbar}\right)^2
\label{eq:scattering-probability-2}
\end{equation}
Equation~(\ref{eq:scattering-probability-2}) is an overestimate,
but probably not very much so.
A derivation of~(\ref{eq:scattering-probability-2})
using standard first-order perturbation
theory is given below in appendix~\ref{a:derivation}.

Consider now
an array of dislocation lines with surface density $n$ and
a wave packet moving in a plane perpendicular to the lines.
As it traverses a distance $L$ it will in general encounter
$L\frac{1}{\Delta k}n$ defects, and the total probability of being
scattered by any of them is
\begin{equation}
P_\subs{scatter along $L$}\sim\left(L\frac{1}{\Delta k}n\right)(a\Delta k)^{d}
\left(\frac{U}{\Delta k
\frac{\partial {\cal E}_n({\bf k})}{\partial {\bf k}}}\right)^2
(a{\bf k})^d 
\left(\frac{\Delta k}{\bf k}\right)
\label{eq:Total-scattering-along-line}
\end{equation}
We want this probability to be much less than one.
A surface dislocation density of $n$ leads to a crystal torsion field
of strength $na$, since each dislocation contributes a Burgers' vector
of length $a$.
The crystal directions can change
appreciably over a length $L$
if the line integral of torsion is of order one, that is $naL\sim 1$.
The two estimates are compatible if
\begin{equation}
(a\Delta k)^{d-1}(a{\bf k})^d
\left(\frac{U}{{\Delta k}
\frac{\partial {\cal E}_n({\bf k})}{\partial {\bf k}}}\right)^2
\left(\frac{\Delta k}{\bf k}\right)
 << 1
\label{eq:Total-scattering-condition}
\end{equation}
We are interested in the case $d=2$.
Assuming
for simplicity
 ${\cal E}_n({\bf k})\sim\frac{1}{2}m_\subs{eff}{\bf k}^2$
we have
\begin{equation}
(a{\bf k})^3
\left(\frac{U}{{\cal E}_n({\bf k})}\right)^2
 << 1\qquad (d=2)
\label{eq:Total-scattering-condition-d-2}
\end{equation}
This bound can evidently only be satisfied, and then
only for sufficiently large crystal momenta
${\bf k}$, if the effective scattering
potential $U$ is much less than the
the largest kinetic energies
in the band.
Condition (\ref{eq:Total-scattering-condition-d-2})
is a serious point.
If $U$ is not sufficiently small then scattering is the
leading effect and considerations pertaining to fine
points of the continuum picture, which we pursue here,
are simply irrelevant to electron motion in quantum dots.
We remark in contrast that in the $3D$ situtation 
(\ref{eq:Total-scattering-condition}) becomes
\begin{equation}
(a\Delta k)(a{\bf k})^4
\left(\frac{U}{{\cal E}_n({\bf k})}\right)^2
 << 1\qquad (d=3)
\label{eq:Total-scattering-condition-d-3}
\end{equation}
which can always be satisfied if $\Delta k$ is small enough.

\section{Possible torsion effects in quantum dots}
\label{s:defect}
I will now consider the following possible types of
effects: i) classical motion;
ii) non-relativistic quantum motion; iii) spin-torsion coupling
terms derivable from the non-relativistic limit of the Dirac
equation minimally coupled to torsion and iv)
phenomenological spin-torsion couplings appropriate for the situation
of quantum dots.

When classical or semi-classical descriptions are valid
the electronic motion in the dot is effectively two-dimensional.
Let the two directions of the dot be $1$ and $2$ and the third
vertical direction $3$.
For points i) and ii)
we only need to consider the possible effects of
elements $S_{121}$ and $S_{122}$ of the crystal torsion matrix,
i.e. only edge dislocations, while for iii) and iv) we also
need to consider screw dislocations.

It is convenient to express torsion in lattice unit $a$.
Kr\"oner discusses a crystal with one
edge dislocation per every ten-by-ten atoms, all Burgers vectors
of all dislocations oriented the same way.
In this situation crystal torsion would be $0.01/a$,
and one would have to
move a distance $100a$ to see
to change the lattice directions change appreciably.
It is not clear if this example is realistic, the
surface defect density being about $10^{17}\hbox{m}^{-2}$.
As the standard set-up I will thus consider one
a dislocation per every hundred-by-hundred atoms.
Torsion would then be $10^{-4}/a$, and one would
have to move a distance of $10^{4}a$ 
 to change the crystal orientations.
Crystals with lower dislocation densities could surely be fabricated,
but then we would need macroscopic quantum dots to see the possible effects
of crystal torsion.

With quantum dots of micron size
the bounds on dispersion from
section~\ref{s:geodesics-vs-autoparallels} can be satisfied.
We would see the lattice orientations change when we move from one
side of the dot to the other, i.e. $L\sim 10^4a$.
If we choose the spread of the wave packet $\Delta k$ to be about 
$10^{-3}/a$. then $1/\Delta k << L << L_\subs{disp}$ with an order
of magnitude at the lower inequality and two orders of magnitude at
the upper inequality. 
The semi-classical argument predicts that wave packets
follow classical trajectories in a Hamiltonian given by the
band energy as functions of crystal momentum. This structure stems
from solving for the Bloch states in the crystal potential,
an information which
is not contained in the continuum description of the crystal
as a metric space with curvature and torsion.
In the general case neither geodesics nor autoparallels in crystal space
therefore have any particular relevance to the problem.
In the special case of a spherically symmetric band structure we do
however recover motion along straight lines, i.e. geodesics
in crystal space.  
Motion along autoparallels would on the other hand
be on curved paths as in the
example discussed in section~\ref{s:classical-quantum}.
If the approximation of a spherically symmetric band structure is
a sufficiently good one, and
if the local scattering potentials are weak enough that the
wave packet does not loose coherence, then the difference
between geodesics and autoparallels could in principle
be observable.

The scale separation is even
more favorable for a description in terms 
of a non-relativistic quantum particle.
Consider the ground state or a low-lying excited
state in the example discussed before.
If the prescription (\ref{eq:Laplace-Kleinert})
were correct, they would be solutions to
the eigenvalue equation
\begin{equation}
\left(-\frac{\hbar^2}{2m_\subs{eff}}\nabla^2 + \sum_i U(x-x_i)
+\frac{\hbar^2}{m_\subs{eff}} S^l\partial_l
\right)\psi = E\psi
\label{eq:Kleinert-eigenvalue}
\end{equation}
where $U$ is the scattering potential from dislocations
at points $x_i$
and $S^l$ is the crystal torsion vector,
which in the example under discussion would
be about $10^{-4}/a$.
The derivative
$\partial_l$ acting on a low-lying state is also about
$10^{-4}/a$ if the
lateral size
of the dot is in the micron range.
Hence the perturbation $S^l\partial_l$ 
is of order $10^{-8}/a^2$, which is comparable to the energy gap
between the ground state and the low-lying states
of (\ref{eq:Kleinert-eigenvalue}).
As discussed in above in section~\ref{s:classical-quantum}
the perturbation
breakes 
hermiticity, and hence leads, if  large enough, to
complex eigenvalues, i.e. states exponentially decaying or
growing in time. 
On a conceptual level the problem is
primarly the exponentially growing modes.
One might imagine that the defects 
could perhaps
excite certain electronic states in the dot, and that for
some transient time the growth rates
 of those states could be decribed
by the eigenvalues of (\ref{eq:Kleinert-eigenvalue}).
This picture does  not look physically
very plausible, and could probably be
ruled out by experiments, at least on a qualitative level.

Let us now turn to possible spin-torsion couplings.
The most straight-forward, but also certainly
the smallest, are the non-relativistic
limits of the vector and axial vector
 couplings in the Dirac equation, 
(\ref{eq:Zeeman-coupling}) and (\ref{eq:Lammerzahl-coupling}).
In the first case we have the curl of the torsion vector.
Let us assume that the torsion varies in a controlled manner
in the plane of the dot.
We only have to consider edge dislocations, since the vector
$\vec S$ vanishes for screw dislocation.
When the density of edge dislocations varies normally
to the vector $\vec S$ we would have an effective coupling
analogous to the Zeeman-effect, 
with an effective
 magnetic field pointing in the vertical direction
(out of the plane of the dot). 
In the axial vector case
we only have to consider screw dislocations,
$\tilde S_0$ being the sum of densities of screw dislocations
in all directions.

We can also discuss more phenomenological spin-torsion
coupling terms.
In the physical situation of a quantum dot 
the vertical direction is determined 
by the gradient of dopant concentration.
We can therefore form a combination of torsion and the vertical
direction vector which transforms as an axial vector,
namely $\hat S^l = \epsilon_{i}\,^{jk} \hat n_3^i S_{jk}\,^{l}$.
This is just the vector $\vec b$ which results from circling
a dislocation in the $12$-plane, i.e. $\hat S$ could have
elements both in and out of the plane.
We then consider a term like 
$\kappa \hat S \cdot \vec\sigma$.
Torsion has  physical dimension of inverse length,
hence $\kappa$ should have dimension
$\hbox{mass}\cdot(\hbox{length})^3(\hbox{time})^{-2}$.
The only combination of
Planck's constant $\hbar$, the lattice constant $a$
and the effective
electron mass $m_\subs{eff}$, which has this dimension is
${  {\hbar^2 }\over{a m_\subs{eff}} }$.
This coupling can be compared dimensionally with
(\ref{eq:Zeeman-coupling})
and (\ref{eq:Lammerzahl-coupling}), which have the bare
electron mass instead of $m_\subs{eff}$ and a space derivative
instead of $1/a$.

The energy splitting of two states with spin respectively
parallel and antiparallel to $\hat S$ will then be 
will then be $\kappa$ times the average value of
$\hat S$ in lattice units.
With the standard case of one dislocation per hundred-per-hundred atoms
this gives
\begin{equation}
\Delta E_\subs{torsion} \approx \frac{\kappa}{ 10^{4}a}
\end{equation}
The frequency of the transition is hence 
\begin{equation}
\nu_\subs{torsion} \sim  10^{-4} { {\hbar}\over{m_\subs{eff} a^2} }
\approx 100 \mbox{Hz}
\end{equation}
where I have for simplicity  estimated $m_\subs{eff}$ with the
electron rest mass and $a$ to be 1 \AA .
The two terms derivable from the Dirac equation discussed above
are even smaller,
approximately by a factor $a/L$.

\section{ Conclusions}
\label{s:conclusion}

I have in this paper taken up a suggestion of Kleinert that the
motion of a classical of a quantum particle in a space with torsion
could be relevant to describe some properties of crystal with defects.
I have focused on the example of electron motion in quantum dots,
since these systems most nearly realise the textbook example of
electron motion in potential wells.

In a space with torsion there are two different priviliged classical motions,
geodesics and autoparallels. 
A semi-classical analysis carried out in
section~\ref{s:geodesics-vs-autoparallels}
indicates that at some scales of space and time geodesic motion
could be observed, if the band energy structure is spherically
symmetric, and the local scattering potential from dislocations
sufficiently weak.
The last condition is very stringent in $2D$ geometries such as
quantum dots, and local scattering is in fact likely to be the
leading effect. A continuum description of the motion of an electron
in a crystal with dislocations is therefore probably not 
very
useful in $2D$.
Assuming nevertheless, for the sake of the argument,
that the local scattering potential is weak,
the conditions for the distinction between geodesics and autoparallels
could probably be experimentally realised. These issues
are discussed in section~\ref{s:defect}.
If the band  energy structure is not spherically symmetric the
problem is more
complicated and neither geodesics nor autoparallels would 
follow from the semi-classical argument.

The distinction between 
geodesics and autoparallels
carry over to the quantum mechanics
of a non-relativistic particle, i.e. to the proper generalization of
the Schr\"odinger equation to a space with torsion.
The analogy with geodesics would be the Laplace-Beltrami operator,
while the analogy with autoparallels would be another
differential operator introduced by Kleinert.
The difference between the two operators can be relatively
large, and hence observable.
Since it breakes hermiticity.
it leads to modes 
exponentially growing or
decaying in time,
an effect which should be
falsifiable by experiments.

The spin-torsion coupling terms, both the ones derivable from the
Dirac equation and more phenomenological ones, are all very small
and give transitions with frequencies less than
1~kHz, or even less than that.

Hence we conclude 
that whatever the motion of a classical or quantum
particle in a space with torsion should be, 
the issue has little bearing
on electron motion in quantum dots.
The semi-classical motion of wave packets can,
in favorable cases, be arranged to be
along geodesics, i.e. along straight lines, but autoparallels
are never relevant.
Bound electronic states in a quantum dot may to some accuracy be
described as the eigenstates of the Laplacean
with Dirichlet boundary conditions,
but the alternative introduced by Kleinert has eigenstates that are
exponentially growing or decaying in time, and this contradicts
the very notion of stationary bound states in the dot.
The spin-coupling terms are not in contradiction with theoretical arguments,
nor with concievable experiments. But their effects are so weak it would
be very difficult to observe them.

\section{Acknowledgments}
I thank Nuno Barros e S\'a, 
Ingemar Bengtsson, Alexander Grishin,
Dmitri Leites,
Paolo Muratore Ginanneschi,
Arne Nordmark,  and
Per Salomonson for discussions.
This work was supported by the Swedish Natural Science
Research Council through grant S-FO-1778-302,

\appendix

\section{Computation of the wave packet-defect cross-section}
\label{a:derivation}

The wave packet interacts with the defect
during a time of about
\begin{equation}
T_\subs{scatter}=\frac{\hbar}
{\Delta k|\frac{\partial{\cal E}_n({\bf k})}{\partial{\bf k}}|}
\label{eq:T-scatter-definition}
\end{equation}
It is therefore convenient to regard 
the defect as a time-dependent perturbation,
switched on 
before  $-T_\subs{scatter}$ and switched off 
after $T_\subs{scatter}$.
For definitiveness take
$V(x,t) = V(x) \exp(-\frac{1}{2\sigma^2}(t/T_\subs{scatter})^2)$
with $\sigma$ some number which will eventually be taken
to infinity.

Let us take a lattice of large finite size $\Lambda$ with the
isolated defect centered at the origin. The values of the wave vector
$k$ then come in integer multiples of $2\pi/\Lambda$.
The time-dependent Bloch states are denoted $\Psi_{n,k}^{(0)}(x)$,
and are normalized by $\int_{\Lambda^d}|\Psi_{n,k}^{(0)}(x)|^2 =1$.
The wave packet formed by Bloch waves
from the $n$'th band is given by $\sum_{k} \Psi_{n,k}^{(0)} a_{k}^{(0)}$,
where the amplitudes are normalized by $\sum_{k} |a_{k}^{(0)}|^2 =1 $.

When the wave packet has passed over the defect,
i.e., after the perturbation has been switched off,
we  have to first order in the perturbation the out state
\begin{equation}
\Psi(\hbox{out}) = 
\sum_k \Psi_{n,k}^{(0)} \left(a_k^{(0)}+\sum_l U_{kl}a_l^{(0)} \right)
\qquad U_{kl} = -\frac{i}{\hbar}
\int_{-\infty}^{\infty} V_{kl}(t) e^{i\frac{E_k-E_l}{\hbar}t}
\label{eq:out-state}
\end{equation}
where $V_{kl}(t)$ is the time-dependent matrix element between the
time-independent Bloch states $k$ and $l$. The time dependence
of the Bloch states have been written out explicitly in the
matrix element in (\ref{eq:out-state}).
The difference between $\Psi(\hbox{out})$ and the unperturbed
wave packet at the same time is
$\delta\Psi(\hbox{out}) = \sum_{kl} \Psi_{n,k}^{(0)} U_{kl}a_l^{(0)}$.
It is orthogonal to the unperturbed state.
The absolute square of $\delta\Psi$ hence represents the probability
to be in an out state orthogonal to the unperturbed wave packet.

I disregard scattering out of the band. Then,
for pairs of states such that $|E_k-E_l|>> \frac{\hbar}{T_\subs{scatter}}$
the transition probability amplitude $U_{kl}$ will be very small.
Inserting the definition of $T_\subs{scatter}$ this means that only
wave vectors in a thin shell of thickness about $\Delta k$ 
around the surface given by $E_k=E_{\bf k}$ are scattered.
The matrix element is 
\begin{equation}
V_{kl}(t) = <k|V|l>
\exp(-\frac{1}{2\sigma^2}(t/T_\subs{scatter})^2)
\label{eq:matrix-element-estimates}
\end{equation}
where $<k|V|l> = \int(\Psi_{n,k}^{(0)})^* (x) V(x)
\Psi_{n,l}^{(0)}(x)\, dx$.
If both wave vector are within  
$\Delta k$ of ${\bf k}$ then, because $1/\Delta k$ is assumed much larger than $a$,
\begin{equation}
<k|V|l> =
\left(\frac{a}{\Lambda}\right)^d U \qquad |k-l|\sim\Delta k
\label{eq:matrix-element-estimates-special-case}
\end{equation}
where $U$ is some measure of the strength of the potential.
In general, if $|l>$ is in the wave packet we can always 
for the same reason substitute $<k|V|l>$ with $<k|V|{\bf k}>$.
We have now
\begin{equation}
U_{kl} = -\frac{i}{\hbar}<k|V(x)|l>
\exp\left(-\frac{1}{2}\left(\sigma T_\subs{scatter}
(\frac{E_k-E_l}{\hbar})\right)^2\right)
\sqrt{2\pi\sigma^2 T_\subs{scatter}^2}
\label{eq:transition-amplitudes-estimates}
\end{equation}
The total probability of scattering is
\begin{equation}
|\delta\Psi|^2 = \sum_{nkl}U_{nk}^* U_{nl}( a_k^{(0)})^* a_l^{(0)}
\label{eq:total-squared}
\end{equation}
Let the amplitudes be given as $a_k^{(0)} = \frac{(2\pi)^{d/2}}
{\Lambda^{d/2}}f_0(k)$,
where the smooth function $f_0(k)$ is normalized by $\int_{\Lambda^d}
|f_0(k)|^2 =1$. For definitiveness take $f_0(k)$ to be a Gaussian
centered at ${\bf k}$ with width $\Delta k$:
\begin{equation}
f_0(k) = \exp\left(-\frac{1}{2}\left(\frac{k-{\bf k}}{\Delta k}\right)^2
\right)\left(\pi(\Delta k)^2\right)^{-d/4}
\end{equation}

We now wish to compute $\sum_{l}U_{nl}a_l^{(0)}$
where $n$ is anywhere on the shell of scattered wave vectors.
We have to separate components of wave vector $l$ parallel
and orthogonal to 
$\frac{\partial{\cal E}_n({\bf k})}{\partial{\bf k}}$.
By expanding 
\begin{equation}
E_n-E_l = (E_n-E_{\bf k}) -  \frac{\partial{\cal E}_n({\bf k})}{\partial{\bf k}}
(l_{\parallel} - {\bf k}_{\parallel})
\end{equation}
and using the definition of $T_\subs{scatter}$
from (\ref{eq:T-scatter-definition})
we  have 
\begin{eqnarray}
\sum_{l} U_{nl} a_l^{(0)}
&=& -\frac{i}{\hbar}<n|V(x)|{\bf k}>
\sqrt{2\pi\sigma^2 T_\subs{scatter}^2}
\left(\frac{\Lambda}{2\pi}\right)^{d/2} 
\left(\pi(\Delta k)^2\right)^{-d/4}
\left(2\pi(\Delta k)^2)\right)^{d/2}\nonumber \\
&&
\sqrt{\frac{1}{\sigma^2+1}}
\exp\left(-\frac{1}{2}\frac{\sigma^2}{\sigma^2+1}
\left(\frac{T_\subs{scatter}(E_n-E_{\bf k})}{\hbar}\right)^2\right)
\end{eqnarray}
The contribution to (\ref{eq:total-squared}) from a given wave vector
$n$ can therefore be written
\begin{eqnarray}
|\sum_{l} U_{nl}  a_l^{(0)}|^2  &=&
\left(\frac{|<n|V|{\bf k}>| T_\subs{scatter}}{\hbar}\right)^2 
\frac{\sigma^2}{\sigma^2+1}
(\Delta k)^{d}
(2\pi)^{d/2+1} 2^{d/2} \nonumber \\
&& \left(\frac{\Lambda}{2\pi}\right)^d
\exp\left(-\frac{1}{2}\frac{2\sigma^2}{\sigma^2+1}
\left(\frac{T_\subs{scatter}(E_n-E_{\bf k})}{\hbar}\right)^2\right)
\label{eq:contribution-from-n}
\end{eqnarray}

When we sum over $n$ we can take out a value ${\bf n}$ such that
$E_{\bf n}=E_{\bf k}$, and then integrate parallel to
$\frac{\partial E}{\partial {\bf n}}$.
That will give a term
$\sqrt{\frac{2\pi(\sigma^2+1)\hbar^2}{2\sigma^2T_\subs{scatter}^2
(\frac{\partial E}{\partial {\bf n}})^2}}$.
Incorporating the partial derivative in a delta function we can hence write

\begin{eqnarray}
|\delta\Psi|^2 &=&
\sqrt{\frac{\sigma^2}{\sigma^2+1} }
(\Delta k)^{d+1} (2\pi)^{d/2+2} 2^{d/2-1}
\left(\frac{T_\subs{scatter}}{\hbar}\right)^2
|\frac{\partial E}{\partial {\bf k}}|
\nonumber \\
&& \left(\frac{\Lambda}{2\pi}\right)^{2d}
\int  \delta(E_n-E_{\bf k}) |<n|V|{\bf k}>|^2\, dn
\label{eq:total-squared-final}
\end{eqnarray}
Here we can take the limit of $\sigma$ going to infinity
and we have thus removed the spurious time-dependence of the 
perturbation.

By order-of-magnitude estimate we take
 $|<n|V|{\bf k}>|= U\left(\frac{a}{\Lambda}\right)^d$
for all states $n$ on the energy shell given by
$E_n=E_{\bf k}$.
We can then take the matrix element
outside the integral
$\int \delta(E_n-E_{\bf k})\, dn$, which is
the density of states at energy $E_{\bf k}$.
If we further assume that $E_n = \frac{1}{2}m_\subs{eff}n^2$,
then the density of states is equal to
$ \frac{\Omega_d {\bf k}^d}{2E_{\bf k}}$,
where $\Omega_d$ is the area of the unit sphere in $d$ dimensions.
We can finally take out from
(\ref{eq:total-squared-final}) the combination
$\Delta k |\frac{\partial E}{\partial {\bf k}}|\frac{1}{2E_{\bf k}}$
which can be rewritten more simply as
$\frac{\Delta k}{\bf k}$,
and this gives (\ref{eq:scattering-probability-2}).

If we assume instead that matrix element is only non-zero between
states in the wave packet, then we can 
estimate the integral in (\ref{eq:total-squared-final}) by
$\int_{|n-{\bf k}|\leq \Delta k} \delta(E_n-E_{\bf k})\, dn$
which is about $(\Delta k)^{d-1}/|\frac{\partial E}{\partial {\bf k}}|$. 
The  probability of being scattered from one wave packet to
another is hence
\begin{equation}
P_\subs{scatter within $\Delta k$} \sim \left(a\Delta k\right)^{2d} 
\left(\frac{UT_\subs{scatter}}{\hbar}\right)^2
\label{eq:scattering-probability-1}
\end{equation}
which can also be derived by straight-forward
dimensional analysis,
since the probability density of the wave packet is about $(\Delta k)^{2d}$.


\begin{references} 

\bibitem{KirkReed} {\it Nanostructures and Mesoscopic Systems},
W.P.~Kirk and M.A.~Reed, eds., Academic Press San Diego (1992).
\bibitem{Kastner}
M.A.~Kastner, {\it Rev. Mod. Phys.} {\bf 64} (1992), 849.

\bibitem{BarangerJalabertStone}
H.U.~Baranger, R.A.~Jalabert and A.D.~Stone,
{\it Chaos} {\bf 3} (1993), 665.

\bibitem{TomsovicHeller}
S.~Tomsovic and E.~Heller,
{\it Phys. Rev. Lett.} {\bf 70} (1993), 1405.

\bibitem{Marcus}
C.M.~Marcus, S.R.~Patel, A.G.~Huibers, S.M.~Cronenwett, M.~Switkes,
I.H.~Chan, R.M.~Clarke, J.A.~Folk, S.F.~Godijn,
K.~Chapman and A.C.~Gossard,
{\it Chaos, Solitons \& Fractals} {\bf 8} (1997), 1261, 

\bibitem{Mathur}H.~Mathur et al, Phys. Rev. Lett.
{\bf 74} (1995), 1855.

\bibitem{Agam}
O.~Agam, N.~Wingreen, B.~Altshuler, D.C.~Ralph and M.~Tinkham,
``Chaos, interactions and nonequilibrium effects in the
tunneling spectra of small metallic particles'',
cond-mat/9611115.

\bibitem{Bruus}
H.~Bruus, C.H.~Lewenkopf and E.R.~Mucciolo,
{\it Phys. Rev. B} {\bf 53} (1996), 9868.


\bibitem{Kleinert} 
	H. Kleinert,
{\it Path Integrals in Quantum Mechanics, Statistics and Polymer Physics},
World Scientific, Singapore 1995.

\bibitem{KleinertFiziev}
H. Kleinert and P. Fiziev, {\it Europhys. Lett.} {\bf 35} (1996), 241.

\bibitem{KleinertLANL}
H. Kleinert and S.V. Shabanov,
``Spaces with torsion from embedding and the special
role of autoparallel trajectories'',
gr-qc/9709067;
H. Kleinert, ``Quantum equivalence
principle'', quant-ph/9612040;
H. Kleinert, ``Nonholonomic mapping principle for
classical and quantum mechanics in spaces with
curvature and torsion'',
FU-Berlin preprint (1997) and {\it Acta Phys. Pol.} (in press);
P. Fiziev and H. Kleinert, ``Anholonomic transformations of Mechanical
action principle'', gr-qc/9605028;
H. Kleinert and A. Pelster, ``Lagrange mechanics in spaces with curvature
and torsion'', gr-qc/9605046.

\bibitem{Hehl}
F.W.~Hehl, P.~van der Heyde and G.D. Kerlick,
{\it Rev. Mod. Phys.} {\bf 48} (1976), 393.

\bibitem{Schrodinger}
E.~Schr\"odinger, {\it Space-time structures},
Cambridge University, Cambridge (1960).

\bibitem{Schouten}
J.A.~Schouten, {\it Ricci calculus},
Springer, Berlin (1954).



\bibitem{Nuno}
N. Barros e S\'a, ``Autoparallels from a variational
principle'', Stockholm University preprint, October 1997.

\bibitem{ArnoldKozlovNeishtadt}
V.I. Arnold, V,V, Kozlov and A.I. Neishtadt,
{\it Dynamical Systems III}, Springer-Verlag, Berlin (1988).


\bibitem{Poincare}
H. Poincar\'e, C. R. Acad. Sci. (Paris) {\bf 132} (1901),
369-371.


\bibitem{Bondi}
H.~Bondi,
{\it Proc. Roy. Soc. Lond. A} {\bf 405} (1986), 265.

\bibitem{GarciaHubbard}
A. Garcia and M. Hubbard
{\it Proc. Roy. Soc. Lond. A} {\bf 418} (1988), 165.

\bibitem{Nordmark}
A.~Nordmark, H.~Ess\'en and M~Lesser, 
``Unidirectional spin and nonholonomic dynamics'',
in {\it Proceedings}, 3rd EUROMECH Solid Mechanics Conference, 
Stockholm August 18-22,
1997.


\bibitem{Weinberg}
S.~Weinberg, {\it Gravitation and Cosmology},
John Wiley \& Sons (1972).

\bibitem{BerlineGetzlerVergne}
N.~Berline, E.~Getzler and M.~Vergne,
 {\it Heat Kernals and Dirac Operators},
Springer Verlag, Berlin (1991).



\bibitem{Audretsch}
J.~Audretsch, {\it Phys. Rev.} {\bf D 24} (1981), 1470.

\bibitem{Lammerzahl}
C.~L\"ammerzahl, ``Constraints on space-time torsion from
Hughes-Drever experiments'', gr-qc/9704047.


\bibitem{HehlDatta}
F.-W.~Hehl and B.K.~Datta, {\it J. Math. Phys.}
{\bf 12} (1971), 1334.

\bibitem{Kroner} 
        E. Kr\"oner,
{\it Continuum Theory of Defects}, in {\it Physics of Defects} R.~Balian ed.,
Les Houches XXXV, North-Holland (1981).

\bibitem{AshcroftMermin} 
N.W.~Ashcroft and N.D.~Mermin,
 {\it Solid State Physics},
Sauders College Publishing (1976).

\bibitem{Kittel} 
C.~Kittel,,
 {\it Introduction to Solid State Physics,
5th Edition}
John Wiley \& Sons, New York (1976).



\end{references}
\end{document}